# Distributed High Speed Optical Network for Digital Radar Systems

Vishal Maheshwari, K.Sreenivasulu, Mohit Kumar, Dr. Vengada Rajan, Sumant Pal, Mohana Kumari
Electronics and Radar Development Establishment (LRDE)
C.V. Raman Nagar, Bangalore - 560 093, INDIA
vishalmaheshwari1990@gmail.com

*Abstract* — Modern Digital radar systems with multiple digital beam forming capability are built of large number of receivers and requires high speed data interface links for transmission of receiver baseband data to processor units. High data throughput (>250Mbyte/sec) from typical eight channel receivers will be transmitted to Digital beam former over high-speed serial interface links over optical channel. Currently for digital radar systems with sub array level beam former distribution of receiver data is through point to point optical interface links. For the modern element level digital beam forming radars the distribution of baseband data increases the design complexity. In this paper novel scheme of usage of distributed optical interface network is discussed using high-speed optical transport networks, FPGAs as well as distributed techniques to over the above problem. The recent advances in optical communication and feature of FPGA devices are utilized in implementation of optical distribution networks and these schemes are covered in this paper.

Key words: optical transport networks (OTNs), FPGAs, Digital Beam Formers, Digital Radars

## I  INTRODUCTION

Communication is becoming a crucial area to look into for providing high speed data interface links for fast and secure transmission. Conventional radar system has lots of RF cables along with its complex connectivity. So weight, power, distortion of signals, EMI/EMC problems, less bandwidth has narrowed down secure and fast digital and RF transmission. To make radar communication more efficient, receive data has to be processed as fast as possible and appropriate control commands to be given to antenna units to perform all necessary actions. This is possible using optical interconnects in radar system because of its high bandwidth, less losses, no EMI/EMC problem, less weight and less power consumption.

There are many RF, digital and control signals in radar exciter and receiver unit like local oscillators, RF transmit energy, clocks etc which can be converted to optical domain, multiplexed and send over a single fiber. This paper gives detailed idea of implementation of optical technology in radar communication system. Paper highlights the use of Optical transmitter chip used in radar exciter to convert all RF signals to optical signals and method to pass all optical signals to all Transmit Receive Modules (TRMs) used in Antenna unit through fiber optic junction box (FOJB). Similarly during receive path, all RF signals are digitalized and converted into optical signals which are fed to Digital Beam Former Unit (DBFU) through digitalized optical fiber channel via FOJB and thus further processing is done at DBFU. Some optical commands are also sent from DBFU through optical uplink like phase gradient value to program the T/R Modules for a particular look angle.

This paper gives overall picture of this processing and provides comparative advantage from conventional radar system in terms of size, power and weight reduction. Paper shows internal design of FOJB, Optical Transmitter Chip and Optical Receiver Chip considering 4 analog and 4 digital channels which can meet our requirement.

## II  SCHEMATIC OF RADAR COMMUNICATION SYSTEM USING OPTICAL SYSTEM

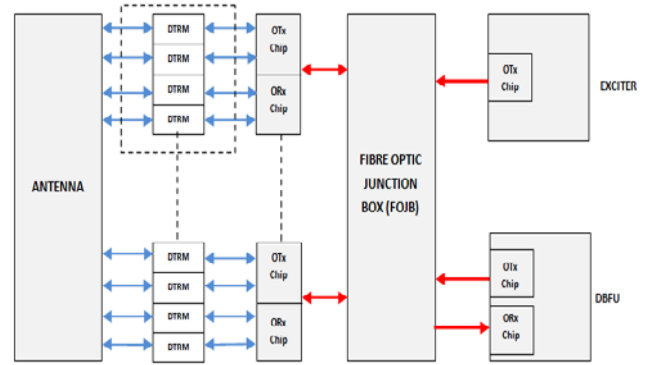

Figure 1. Schematic of Radar communication system using optical Transmitter and Receiver chip

Digital Transmit Receive Module (DTRM) has upconverter path and receive front end, thus it requires LO and Transmit drive signals. It also has Analog to Digital Converter (ADC) chip to digitize receive signals using ADC clock.

*1. Transmission chain*

During transmission period, all RF signals like Local Oscillators (LOs), clocks, transmit drive signals generated by exciter, are fed to Optical Transmitter Chip (OTxC) which is mounted on RF exciter unit as shown in figure 1. OTxC converts RF/digital signals to optical signals and multiplexes them to send over single fiber using Wavelength Division Multiplex (WD Mux) technique. Further this multiplexed optical signal splits into N signals (considering N number of DTRMs) at fiber optic junction box (FOJB) which has 1:N splitter and optical amplifier named as Erbium Doped Fiber Amplifier (EDFA). These signals are fed to ORx chip, where optical signals are demultiplexed using Wavelength Division Demultiplex (WD Demux) technique and converted back to RF signals. RF transmit signal is up converted in DTRM unit and sent to antenna unit for transmission.

*2. Receive chain*

During receive period, 4 RF outputs from antenna unit are collected and down converted to baseband signal and digitalized using ADC in DTRM block. These digitalized signals are converted to optical signals and multiplexed using WDM technique in digital OTxC. Thus multiplexed signals are sent to DBFU through single optical fiber, where it is again de-multiplexed and converted back to digital signal for further processing using digital ORxC.

One group of 4 RF signals is highlighted in rectangular box as shown in figure 1. There are N/4 numbers of such groups. And each group undergoes through same process as explained above.

## III TRANSMITTER RECEIVER CHIP

*1. Transmitter chip*

Transmitter chip is used to send RF/Digital signal over single optical fiber using WDM technique after converting it to optical signal. It consists of many lasers of different frequencies and optical modulator along with optical signal combiner (WD Mux), EDFA (optical amplifier) as shown in figure 2. Number of lasers is dependent on number of signals need to be multiplexed and required in DTRM. For the purpose of obtaining good SFDR and high signal fidelity, it is required to use very high power and low RIN laser. We are ultimately splitting multiplexed optical signal at FOJB, thus we require EDFA to compensate optical losses.

Our study on size reduction suggests that high power laser diode must be mounted on OTxC with very good conductivity. Here we need to use many lasers for particular signals to multiplex, so precise wavelength is required which we can get using thermal tuning. Hence optical high power laser is packaged with modulator and isolator (to reduce back reflected light).[1]

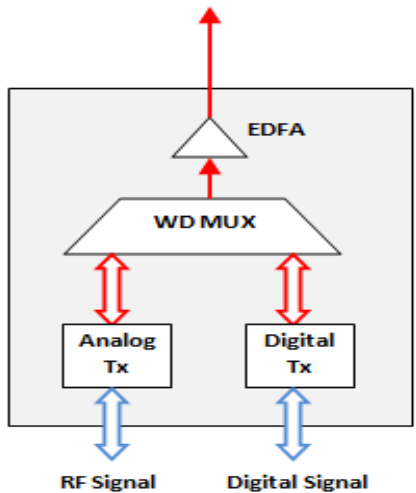

Figure 2. Internal design schematic of Optical Transmitter Chip

*2. Receiver chip*

Receiver chip consists of optical De-multiplexer (WD Demux) and photo detector as shown in figure 3. It demultiplexes optical signal and converts optical signal to RF/digital signal. ORxC needs high saturation detector to detect high power signal also. There we use analog photo detector to detect RF signals and digital photo detector to detect digital signals.

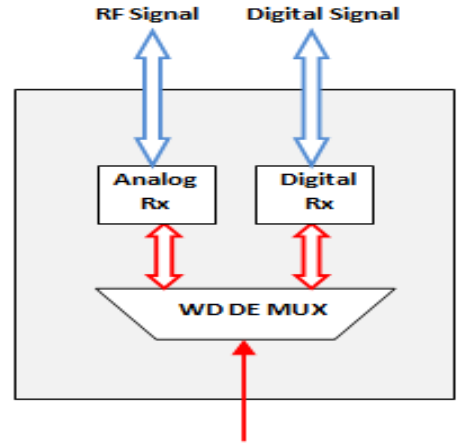

Figure 3. Internal design of Optical Receiver Chip

## IV OUR REQUIREMENTS

Our design for the OTxC and ORxC requires to transmit RF signal frequency of 10MHz to 18 GHz, where receive RF signal has power level of -10 to -130 dBm. For multiplexing optical signal, light source of 1300-1650 nm wavelengths is required. Noise figure should not degrade from input noise figure of analog channels by more than 0.5 to 1dB and phase noise and spurious level shall not degrade by more than 1 to 2 dB during conversion. Apart from these parameters of signals, other performance requirement parameters also to be considered, such as RF cross talk, timing jitter, rise and fall time, pulse skew and signal to noise (SNR) degradation. Study of its preliminary performance guide us to select component and package them carefully. Power, size and weight reduction analysis is carried out later in this paper.[2]

## V ARCHITECTURAL/TECHNOLOGICAL ASPECTS OF DESIGN

Our link design can be implemented using either direct modulation or external modulation scheme in both Volume Bragg Grating (VBG) and Arrayed Waveguide Grating (AWG) based technology. This can be done either using Silicon photonics or (High Integrated Photonic) HIP technology. VBG is used to provide stabilized wavelength of light and WDM multiplexing/ de-multiplexing. VBG also act as an isolator to the laser source. AWG is basically used in WD Mux (Demux) technology. it makes efficient system to multiplex large number of wavelengths into single fiber.

It is better to design chip using athermal silica AWG instead of doing it on III-V HIP technology. Athermal AWG is suited to stabilize wavelength of light without any thermoelectric cooling. Athermal AWG performs well without any yield problem during production. And Silica provides low losses and better

matche to fiber core size. Silicon technology is not supported by VBG based link design.[3]

In the direct modulation scheme, the driving current to a directly modulated (DM) semiconductor laser is varied according to the data to be transmitted. But in the external modulation (EM) scheme, the laser that is subjected to a constant bias current emits a continuous wave (CW) while an external modulator switches the optical power on or off according to the data stream.[4]

Study on design performance of optical chip shows that SFDR can be obtained >55dB for 10MHz bandwidth. NF degradation of <1dB is possible for DM but in case of EM, NF can meet <2dB degradation. Phase noise generally remains below the system noise figure under small signal operation. But during large signal operation, shot noise, power supply noise and other noise play major role to increase system noise. Then noise figure of system is not valid. But for our case, we have considered small signal operation only. Hence phase noise is always lower than system noise figure. Spurs are affected by RF Photonic link which can be controlled by selecting proper components. Rise time and fall time can also be controlled by selecting proper RF optical components such that they provide better bandwidth (should be > 10MHz). Output SNR is equals to the summation of input SNR and NF. Since NF degradation of system is <2dB, hence SNR degradation is also less. RF cross talk comes from ORx Chip and from RF board layout which can be minimized by careful packaging of board. For packaging the analog channels under athermal AWG approach requires more space because EM is required for high power and the temperature control for each laser. So it is better to package these channel separately and combine their output fiber before entering an external AWG to reduce packaging size. It compensates all temperature variation for both analog and digital.[5]

If we categorize our design considering power aspect, then both AWG and VBG based design consume more power in case of EM in comparison with DM. VBG based design consumes less power than AWG based design.

Size and weight analysis of design shows that VBG based design requires less space and has less weight than AWG based approach for III-V HIP technology. But if we consider Silicon (Si) HIP technology, then AWG based approach is best in case of EM and VBG based design is best in the case of DM. Weight and size are more in case of EM rather than DM.

## VI CONCLUSION

Design methodology is more focused on providing excellent performance of optical system in RF links, which is incorporated by using high power and low RIN analog laser along with high saturation current photo detector.Digital link is enhanced by using tunable digital laser. VBG based link design does not support Silicon technology. Majority of power is consumed by analog links but in optical, major power consumer is optical laser and their thermal management. Design can be implemented using HIP technology or Silicon based photonics.VBG is useful to generate light of stabilize wavelength and good WD Mux/ Demux. This concludes that NF (<2dB), Phase noise, SNR degradation, RF crosstalk, SFDR (>55dB @ 10MHz bandwidth), Rise time- fall time and spurs can be controlled. EDFA is used to compensate all optical losses. Both analog and digital signal can be detected from the same design of receiver. Power, size and weight are less for both VBG and AWG based approach in case or direct modulation rather than external modulation. Overall VBG based design approach suits to our requirement.

**References:**

[1] Shuang Liu, Deming Liu, Junqiang Sun, "Multiple-wavelength transmitter for WDM optical network", Frontiers of Optoelectronics in China June 2009, Volume 2, Issue 2, pp 200-203
[2] M. Skolnik, "Introduction to radar systems", McGraw-Hill, 1981.
[3] De-Lu Li, Chun-Sheng Ma, Zheng-Kun Qin, Hai-Ming Zhang, Da-Ming Zhang, Shi-Yong Liu, "Design of athermal arrayed waveguide grating using silica/polymer hybrid materials", Optica Applicata, Vol. XXXVII, No. 3, 2007
[4] Christophe Peucheret and DTU Fotonik, "Direct and external modulation of light", 34129 Experimental Course in Optical Communication, Technical University Denmark
[5] C. Cox, "Analog Optical Link", Cambridge, U.K. : Cambridge University, Press, 2004

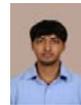
**Vishal Maheshwari** born on 18th December 1990 obtained his B.Tech degree in Electrical Engineering stream from IIT,Gandhinagar in 2012. He is currently working as scientist at Electronics and Radar Development Establishment (LRDE), Bengaluru. Area of specialization is in Digital communication and optical realization

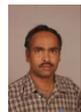
**K Sreenivasulu** received his Diploma in Electronics and Communication Engineering from S.V. Government Polytechnic, Tirupati, Andhra Pradesh in the year 1987. He received his B.Tech degree in Electronics and Communication Engineering from JawaharlalNehruTechnologicalUniversity, Hyderabad in the year 1995. He received M.E. degree in Micro Electronics Systems from Indian Institute Of Science, Bangalore in 2004. He started his professional career as Electronic Assistant in Civil Aviation Department where he worked from 1990 to 1995. Since 1996 he has been working as Scientist in Electronics and Radar Development Establishment (LRDE), Bangalore. His area of work has been design and development of RF and Microwave sub-systems, Digital Radar system, Beam Steering Controller for Active Aperture Array Radars. His interests include VLSI Systems and Programmable Controllers.

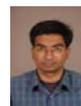
**Mohit Kumar** born on 7thOctober 1980 obtained B.Tech degree in Electronics and Communication from NIT,Jalandhar in 2002. He has completed his MTech from IIT Delhi in Communication Engineering in 2010. He is currently working as scientist at Electronics and Radar Development Establishment (LRDE), Bengaluru. Area of specialization is in Digital Radar Receiver design.